\documentstyle[aps,multicol,epsf]{revtex}
\begin{document}
\title{Stability analysis of the $D-$dimensional nonlinear
Schr\"odinger equation with trap and two- and three-body interactions}
\author{A. Gammal$^{1}$, T. Frederico$^{2}$, Lauro Tomio$^{1}$,
and F.Kh. Abdullaev$^{1}$ \thanks{On leaving from 
Physical Technical Institute, Tashkent, Uzbekistan.}} 
\address{$^{1}$ Instituto de F\'{\i}sica Te\'{o}rica, 
Universidade Estadual Paulista, \\
01405-900 S\~{a}o Paulo, Brazil \\
$^{2}$Departamento de F\'{\i }sica, Instituto Tecnol\'{o}gico da
Aeron\'{a}utica, \\
Centro T\'{e}cnico Aeroespacial, 12228-900 S\~{a}o Jos\'{e} dos Campos, SP,
Brazil}
\date{\today}
\maketitle
\begin{abstract}
Considering the static solutions of the $D-$dimensional nonlinear
Schr\"odinger equation with trap and attractive two-body interactions,
the existence of stable solutions is limited to a maximum critical number
of particles, when $D\ge 2$. In case $D=2$, we compare the variational
approach with the exact numerical calculations. We show that, the addition
of a positive three-body interaction allows stable solutions beyond the
critical number. In this case, we also introduce a dynamical analysis
of the conditions for the collapse.
\end{abstract}
\begin{multicols}{2}
Recent experiments on Bose Einstein Condensation (BEC)~\cite{exp} 
have brought great attention to its theoretical formulation. 
Atomic traps are effectively described by the Ginzburg-Pitaevskii-Gross 
(GPG) formulation of the nonlinear Schr\"odinger equation
(NLSE)~\cite{gin}, which includes two-body interaction.
When the atoms have negative two-body scattering lengths, a formula
for the critical maximum number of atoms was presented in ref.~\cite{WT}.  
In ref.~\cite{ADV,GFT,GFTC}, the formulation was
extended in order to include the effective potential originated from
the three-body interaction. In this case, in three-dimensions, it was
shown that a kind of first order phase-transition occurs. 
In this connection, as also considered in the motivations given 
in \cite{GFT,GFTC}, it is relevant to observe that recently
it was reported the possibility of altering continuously the two-body 
scattering length, from positive to negative values, by means of an 
external magnetic field~\cite{achange}. Within such perspective, the 
two-body binding energy can be close to zero, and one can approach the 
so-called Efimov limit, which corresponds to an increasing number of 
three-body bound states~\cite{efimov}. Near this limit, nontrivial 
consequences can occur in the dynamics of the condensate, such that
one should also consider three-body effects in the effective 
nonlinear potential.

In the present work, we study the critical number of atoms in
arbitrary $D-$dimensions, using a variational procedure; and also
by an exact numerical approach in the case of dimension $D=2$.
The $D-$dimensional NLSE, with attractive two-body interactions, was
previously analyzed in models of plasma and light waves in nonlinear 
media~\cite{zakharov}. The collapse conditions, in this case, were
investigated without~\cite{weinstein} and with~\cite{TW} the harmonic
potential term. In case of $D=$3, it was shown that a
repulsive nonlinear three-body interaction term can extend
considerably the critical limit for the existence of stable
solutions~\cite{ADV,GFT,GFTC}. 

Motivated by the observed high interest in stable solutions for arbitrary 
$D$, we look for variational solutions in  a few significant
cases ($D=$1,2,4 and 5) not previously considered, when a three-body
interaction term, parametrized by $\lambda_3$, is added to the
effective non-linear interaction that contains a two-body attractive term.
Our analysis also shows that, as in case of $D=3$, a kind of
first-order phase-transition can occur when $D\ge 4$, for certain
cases of $\lambda_3\ge 0$.
In the present paper, we have also considered the approach given in
\cite{Pit1}, in order to study the stability conditions
in the case of arbitrary $D$, when the non-linear interaction
contains two (attractive) and three-body terms.

In order to obtain an analytical approach and verify 
the validity of the variational Ritz method, 
we consider in detail the case of $D=2$, with and without
the three-body term, comparing the variational results with exact
numerical calculations for some relevant physical observables. 
In this case, we also discuss how the method given in \cite{weinstein}
can be extended in order to approach analytically the exact value for
the total energy.

By extending the GPG formalism from three to $D$ dimensions, 
including two~\cite{baym96} and three-body interactions in the 
effective non-linear potential~\cite{GFT}, we obtain 
\begin{eqnarray}
i\hbar\frac{d\psi}{dt} =
\left[-\frac{\hbar^2}{2m}\nabla^{2}+\frac{m\omega^2 r^2}{2} 
+\lambda_2|\psi|^2+\lambda_3|\psi|^4
\right]\psi\label{1} ,
\end{eqnarray}  
where $\psi\equiv\psi(\vec{r},t)$ is the wave-function normalized to the
number of atoms $N$, $\omega$ is the frequency of the trap harmonic
potential and $m$ is the mass of the atom.
$\lambda_2$ and $\lambda_3$ are, respectively, the
strength of the two- and  three-body effective interaction, given
in a $D-$dimensional space. 
$r\equiv|\vec{r}|$ is the hyperradius, such that 
$\vec{r}\equiv\sum_{i=1}^{D}r_i\hat{e}_i$ and  
$\nabla\equiv\sum_{i=1}^{D}\hat{e}_i\frac{\partial}{\partial r_i}$  
($\hat{e}_i$ is the unit vector, with $i=1,2,... D$). 

The stationary solutions for the chemical potential $\mu$ are given
by
\begin{equation} 
i\hbar\frac{d\psi}{dt} = \mu\psi.
\label{1.1}
\end{equation}
Considering the general solution of eq.(\ref{1}), 
${\displaystyle i\hbar\frac{d\psi}{dt} = \frac{\delta{\cal
H}}{\delta\psi^\star}}$,
one can obtain the total energy $E$:
\begin{eqnarray}
E&=&\int d^D\vec{r}\;\;{\cal H} ,\;\;\;{\rm with}\label{Etot}\\
{\cal H}&\equiv&
\frac{\hbar ^{2}}{2m}\left| \nabla \psi
\right|^{2}+\frac{m\omega ^{2}r^{2}}{2} \left| \psi\right|^{2}
+\frac{\lambda_2}{2}|\psi|^{4}+
\frac{\lambda_3}{3}|\psi|^{6}
.  \nonumber
\end{eqnarray}   

Here we consider only attractive two-body interaction, which 
is more interesting in the case of trapped atoms. 
For $D=$3, $\lambda_2 \equiv -4\pi\hbar^2|a|/m$,  where $a$ is
the two-body scattering length and $m$ is the mass of the atom.  
In the case of arbitrary $D$, $\lambda_2$ has 
dimension of energy times $L^D$, where $L$ is a length scale in such
space. However, a convenient redefinition of the wave-function in terms of 
dimensionless variables will absorb this constant, as will be shown.  

Our study will be concentrated on the ground state for a spherically
symmetric potential.
We first consider the case of $\lambda_3=0$, using a variational
procedure, with a trial Gaussian
wave-function for $\psi (\vec{r})$, normalized to $N$, given by
\begin{equation}
\psi_{var}(\vec{r})=\sqrt{N}\left(\frac{1}{\pi\alpha^{2}}\frac{m\omega
}{\hbar }\right)^{{D}/{4}}
\exp {\left[ -\frac{r^{2}}{2\alpha^{2}}\left(
\frac{
m\omega }{\hbar }\right) \right] },  \label{varwf}
\end{equation}
where $\alpha $ is a dimensionless variational parameter. 
From eq.~(\ref{Etot}), the corresponding expression for the total
variational energy can be expressed as 
\begin{eqnarray}                        
E_{var}&=&\hbar \omega \frac{N}{\nu}
{\cal E}_{var},\label{EvarND}\\
{\cal E}_{var}&\equiv& \nu
\left( \frac{D}{4\alpha^2}+\frac{D\alpha^2}{4}\right)-
\frac{\nu^2\Omega_D}{4(2\pi)^{D/2}\alpha^D}+ 
\frac{G_3}{6\pi^D}
\frac{\nu^3\Omega_D^2}{3^{D/2}\alpha^{2D}}
,\nonumber\\
\end{eqnarray}                        
where $\Omega_D$ is the solid angle in $D$ dimensions,
\begin{eqnarray}
&&\Omega_D\equiv\frac{2\pi^{D/2}}{\Gamma (D/2)},
\;\;\;\;
G_3\equiv\frac{\lambda_3}{2(\lambda_2)^2}
\hbar \omega \;,
\nonumber\\
&&{\rm and} \;\;\;
\nu\equiv - \frac{N}{\Omega_D}\frac{2\lambda_2}{\hbar\omega}
\left(\frac{m\omega}{\hbar}\right)^{D/2}.
\end{eqnarray}                        
By using dimensionless variables, 
$\vec{x}\equiv \sqrt{{m\omega }/{\hbar }}\; \vec{r}$,  
we redefine the wave-function $\psi$ as 
\begin{equation}
\phi(\vec{x})\equiv
\sqrt{\frac{2|\lambda_2|}{\hbar\omega}} \psi(\vec{r}),
\label{wf}
\end{equation} 
such that
\begin{equation}
\int |\phi (\vec{x})|^2 d^{D}\vec{x} = 
N\left[\frac{2|\lambda_2|}{\hbar\omega}\right]
\left(\frac{m\omega}{\hbar }\right)^{D/2}   
= \nu\Omega_D.
\label{norm}
\end{equation}
The dimensionless equation corresponding to 
eq.~(\ref{1}), can be rewritten as 
\begin{equation}
\left[ \left(- \sum_1^{D}
\frac{d^2}{dx_i^2}+x_i^2\right)-|\phi|^{2}
+G_{3}|\phi|^{4} -2 \beta \right] \phi= 0 ,
\label{schd}
\end{equation}
where $\beta\equiv\mu/(\hbar\omega)$ is the dimensionless chemical 
potential.
From eqs.~(\ref{wf}) and (\ref{varwf}), the trial wave-function can 
be written as
\begin{equation}
\phi_{var}(x)\equiv
\sqrt{\nu\Omega_D}\left(\frac{1}{\pi\alpha^2}\right)^{D/4}
\exp\left(-\frac{x^2}{2\alpha^2}\right),
\label{wfv}
\end{equation}

The variational results, obtained  by using the above expressions can be
extended analytically to non-integer values of the dimension $D$. 
Minimization of the energy [eq. (\ref{EvarND})], with respect to
$\alpha^2$, is done numerically by sweeping over $\alpha^2$ values.
The results for the energy and the chemical potential are shown in Fig. 1.
For each value of $D$, one can observe a critical
number of atoms, $N_c$, related to the critical parameter $\nu_c$, only
when $D\le 2$. This critical limit corresponds to the cusps in the upper
plot of Fig.1 and is also observed using exact numerical calculation for
$D=$3.
It is also interesting to note that for $D>2$ there are two
branches of solutions for ${\cal E}_{var}$ and $\beta$,
one stable and the other unstable. 
In the energy, the lower branch corresponds to stable solutions
(minima), while the upper one gives unstable solutions (maxima).

%%%%%%%%%%%%%%%%%%%%%%%%%%%%%% FIG.1 %%%%%%%%%%%%%%%%%%%%%%%%%%%%%%%%%%%%%%
\begin{figure}[tbp]
\setlength{\epsfxsize}{0.9\hsize} 
\setlength{\epsfysize}{0.8\hsize} 
\centerline{\epsfbox{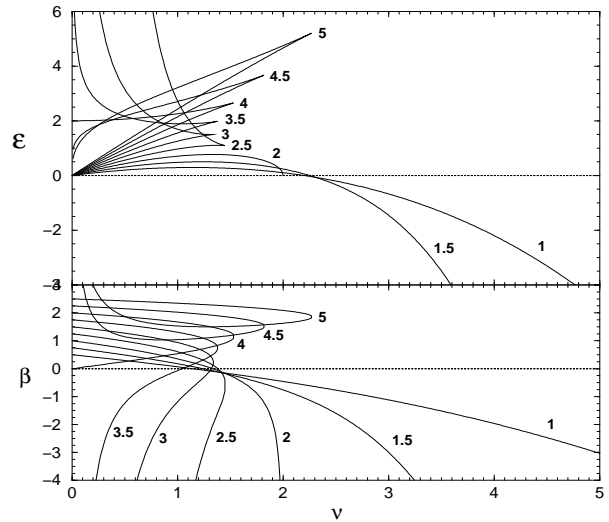}} 
\caption{
The variational energy ${\cal E}$ (upper part) and chemical
potential $\beta$ (lower part), as functions of the reduced 
number of atoms $\nu$, for several values of the dimension
$D$, indicated in each plot. All the quantities are in 
dimensionless units, as defined in the text.
}
\end{figure}
%%%%%%%%%%%%%%%%%%%%%%%%%%%%%%%%%%%%%%%%%%%%%%%%%%%%%%%%%%%%%%%%%%%%%%%%%%%%

The case with $D=2$ is particularly interesting, as no
unstable solutions exist and there are stable solutions only for
$\nu < 2$, such that $\nu_c =2$.
For $D=2$, the minimization of eq.~(\ref{EvarND}) in respect to 
$\alpha^2$ leads to 
\begin{equation}
{\cal E}_{var}=\nu\sqrt{1-\frac{\nu}{2}}.
\end{equation}
The behavior of $\nu$, and the corresponding critical limits, as one
alters the dimension $D$, has other curious particular results. For
example, the critical limit $\nu_c$ has a minimum for $D=3$ ($\nu_c^{(D)}
\ge \nu_c^{(3)}$ for all $D$).

In conclusion of this part of our work, considering arbitrary 
$D$ with $\lambda_3=0$, there are no stable solutions for eq.~(\ref{1}),
if the wave-function $\phi(x)$, given by eq.~(\ref{norm}), is 
normalized to $\nu > \nu_c$. Fig. 1 shows that this restriction is 
strongest for $D=3$: $\nu_c$ is a minimum when compared with 
$\nu_c$ for $D\ne 3$.
This is a relevant result, considering that $\nu$ is directly 
proportional to the number of atoms.  
Also, it is observed that $\nu_c$ increases very fast for $D>3$.
%%%%%%%%%%%%%%%%%%%%%%%%%%%%%% FIG. 2 %%%%%%%%%%%%%%%%%%%%%%%%%%%%%%%%%%%%%% 
\begin{figure}[tbp]
\setlength{\epsfxsize}{1.0\hsize} 
\setlength{\epsfysize}{1.0\hsize} 
\centerline{\epsfbox{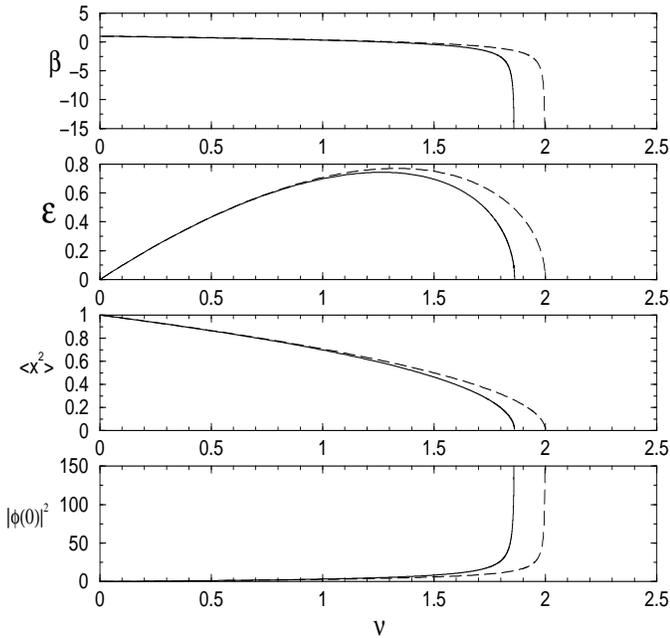}}
\caption{For $D=2$, we present the variational and exact numerical
calculations of the chemical potential ($\beta$), total energy (${\cal
E}$), mean-square-radius ($\langle x^2\rangle$), and central density
($|\phi(0)|^2$), as a function of the reduced number of atoms $\nu$.
All the quantities are in dimensionless units (see text).
The solid line curves correspond to exact numerical results, while the
dashed curves are the variational results.}
\end{figure}
%%%%%%%%%%%%%%%%%%%%%%%%%%%%%%        %%%%%%%%%%%%%%%%%%%%%%%%%%%%%%%%%%%%%%

Next, we also solve equation (\ref{schd}) exactly 
employing the shooting and Runge-Kutta methods, and compare the results 
with the corresponding variational ones. 
In this case, we consider only the particular 
interesting case of $D=2$, with $\lambda_3=0$. The results are shown in 
Fig. 2, for 
the chemical potential $\beta$, the total energy ${\cal{E}}$,
mean-square-radius $\langle x^2\rangle$,
and the central density $|\phi(0)|^2$.  
In order to numerically solve eq.~(\ref{schd}), in the $s-$wave,
we first write it in terms of the single variable 
$x\equiv\sqrt{(x_1^2+x_2^2)}$ and consider the following  
boundary conditions: 
$\phi^\prime(0)=0$ (where $\prime$ stands for the derivative with 
respect to $x$)
and $\phi (x)$ $\to C\exp (-x^{2}/2+[\beta -{1}]\ln (x))$ when
$x\to \infty$, where $C$ is a constant to be determined.
As observed in Fig.2, the critical limit $\nu_c=2$ obtained 
analytically using the variational approach should be compared with
$\nu_c=1.862$, obtained by exact numerical calculation. This critical
limit was first obtained by Weinstein~\cite{weinstein}, in a non-linear
approach with two-body term, without the trapping potential.
The coincidence of the value with our exact calculation is due to the 
fact that at the critical limit the mean square radius goes to zero.

%%%%%%%%%%%%%%%%%%%%%%%%%%%%%% FIG. 3 %%%%%%%%%%%%%%%%%%%%%%%%%%%%%%%%%%%%
\begin{figure}[tbp]
\setlength{\epsfxsize}{0.9\hsize} 
\setlength{\epsfysize}{0.8\hsize} 
\centerline{\epsfbox{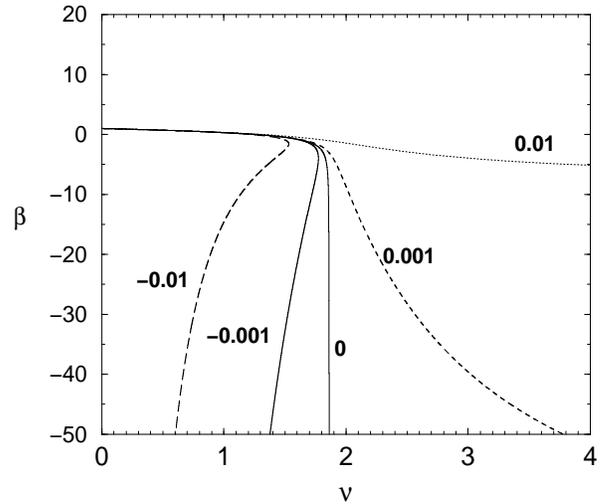}} 
\caption{
Exact numerical solutions for $D=$2, of the chemical potential,
$\beta$, in dimensionless units, given as function of the reduced
number of atoms $\nu$, for different values of the three-body
parameter $G_3$, when the space dimension is $D=$2. 
As shown, only when $G_3\le 0$ the number of
atoms is limited to certain critical number.} 
\end{figure}
%%%%%%%%%%%%%%%%%%%%%%%%%%%%%%        %%%%%%%%%%%%%%%%%%%%%%%%%%%%%%%%%%%%%%
%%%%%%%%%%%%%%%%%%%%%%%%%%%%%% FIG. 4 %%%%%%%%%%%%%%%%%%%%%%%%%%%%%%%%%%%%
\begin{figure}[tbp]
\setlength{\epsfxsize}{0.9\hsize} 
\setlength{\epsfxsize}{0.8\hsize} 
\centerline{\epsfbox{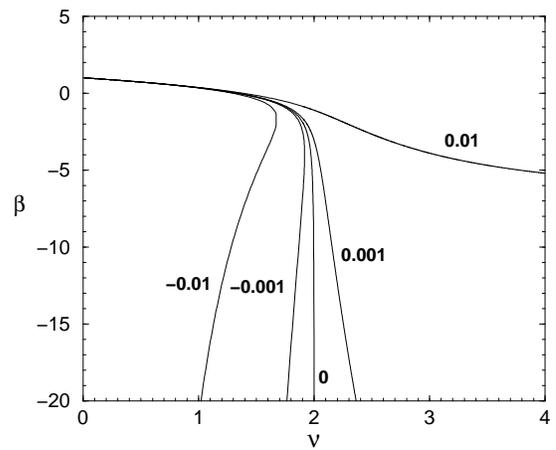}} 
\vskip 0.5cm
\caption{
The variational solutions for $D=$2, of the chemical
potential, corresponding to the exact results given in Fig. 3.} 
\end{figure}
%%%%%%%%%%%%%%%%%%%%%%%%%%%%%%        %%%%%%%%%%%%%%%%%%%%%%%%%%%%%%%%%%%%%%

We have also compared the results obtained by the variational approach
with the exact numerical one, in the case of $D=2$, for several
values
of the three-body interaction term (positive and negative), as shown in
Figs. 3 and 4.  In Fig. 3 we have the exact numerical approach and 
in Fig. 4 we have the corresponding variational results.
By comparing the results we have for $D=2$ (shown in Figs. 3 and 4) with
the ones obtained in ref.~\cite{GFT} for $D=3$, we should observe that no 
first order phase-transition exists in two dimensions. As observed in 
refs.~\cite{ADV,GFTC}, for $D=3$, a first-order phase-transition
can occur in trapped condensed states with negative two-body
scattering length, when a repulsive three-body (quintic) term is added
in the Hamiltonian. As shown in Figs. 3 and 4, with $G_3$ positive the
range of stability for the number of atoms $N$ can be increased
indefinitely; with $G_3$ negative this range is reduced.

 We can analyze the collapse conditions using ``the virial theorem"
approach \cite{Pit1}. The mean square radius,  $\langle r^2\rangle$, of a
$D-$dimensional condensate, is given by  
\begin{eqnarray}
&&\frac{d^2\langle r^2\rangle }{dt^2} + 4\omega^2 \langle r^2\rangle = 
\label{pit}\\
&&\frac{1}{m}\left[{4 \langle H\rangle } +
{\lambda_2}(D-2) 
\langle |\psi|^2 \rangle + 
\frac{4\lambda_3}{3}(D-1)
\langle |\psi|^4 \rangle \right], 
\nonumber\end{eqnarray}
where 
\begin{equation}
\langle {\cal O}\rangle\equiv \frac{1}{N}\int d^D\vec{r}
\psi^\dagger(\vec{r},t)
{\cal O} \psi(\vec{r},t)
\end{equation}
and $\langle H\rangle = E/N$. When $\lambda_3 = 0$ we obtain the equation
derived in \cite{Pit1}. 

We can also write the eq.~(\ref{pit}) in dimensionless units, as
it was done in eqs.~(\ref{wf}-\ref{schd}):
\begin{eqnarray}
&&\frac{d^2\langle x^2\rangle }{d\tau^2} 
+ 4\langle x^2\rangle = \frac{4 {\cal E} }{\nu} + 2f(\tau),
\label{pitx}\end{eqnarray}
where
\begin{eqnarray}
f(\tau)\equiv
\frac{\lambda_2}{|\lambda_2|}\frac{D-2}{4} 
\langle |\phi|^2 \rangle + 
G_3\frac{D-1}{3} 
\langle |\phi|^4 \rangle . 
\label{pitx2}\end{eqnarray}
Using the initial conditions for $\langle x^2 \rangle$ and 
$d\langle x^2 \rangle/d\tau$, where, for simplicity, 
we assume ${d\langle x^2\rangle }/{d\tau} = 0$, 
the solution of eq.~(\ref{pitx}) is given by
\begin{eqnarray}
\langle x^2 \rangle &=& \frac{{\cal E}}{\nu} + 
\left[\left.\langle x^2\rangle\right|_0 -
\frac{{\cal E}}{\nu}\right]\cos(2\tau) 
\nonumber\\
&+&\int_{0}^{\tau}f(\tau')\sin(2(\tau-\tau'))d\tau'.
\label{x}\end{eqnarray} 

The stability regions and the estimates for the collapse time can be
obtained from the analysis of this solution, like as performed for the
case $\lambda_3 = 0$ in \cite{Wad2}.
Let us analyze the dynamics when $D=2$. 
In this case, $\lambda_2$ does not appear explicitly in 
$\langle x^2\rangle$ and $f(\tau)$ also does not depend on this 
parameter:
\begin{enumerate}
\item For a positive $G_3$, negative $\lambda_2$ and ${\cal E}>0$ we
observe that $\langle x^2\rangle$ cannot be zero and the condensate is
stable. The mean square radius of the condensate oscillates in time around
a finite value. This is confirmed by the numerical simulations
(see Figs. 3 and 4).
\item For a negative $G_3$, positive $\lambda_2$ an analysis of
stability like the one performed in ref.~\cite{Wad2} shows that
\newline
{\bf a)} \ When the total energy ${\cal E}<0$, the condensate is unstable
and the wavefields collapse in a finite time at any initial conditions;
\newline
{\bf b)} \ When ${\cal E}>0$, as the function $f(\tau)$ is negative, the 
contribution of the integral term for $\tau<\pi$ is negative. 
Then, we found the collapse condition as 
\begin{equation}
\langle x^2\rangle|_0 \ge 2\frac{\cal E}{\nu}.
\end{equation} 
\end{enumerate}
The same kind of analysis, for $D>2$, is much involved in the present
approach, as the sign of the function $f(\tau)$ is not fixed at opposite
signs for the parameters $\lambda_2$ and $\lambda_3$.

Some information about the dynamics of the collapse can also be obtained
by using the techniques based on integral inequalities~\cite{weinstein,Tur}.
For instance, when $D=2$, we can estimate the three-body term 
contribution in $E$, following the procedure given in \cite{weinstein}
\begin{equation}
\int d^2\vec{r} |\psi|^6 \leq C_2 \left(\int d^2\vec{r}
\frac{|\nabla\psi|^2}{2m} \right)^2
\left(\int d^2\vec{r}|\psi|^2\right) = C_2 K^2 N,
\end{equation}
where $K$ is the kinetic energy and $C_2$ is defined from the 
minimization of the functional
\begin{equation}
{\cal J} = \frac{
\left(\int d^2\vec{r}|\nabla\psi|^2\right)^2
\left(\int d^2\vec{r}|\psi|^2 \right) 
}
{\int d^2\vec{r} |\psi|^6}.
\end{equation}
Combining with the corresponding estimate for $\int d^2\vec{r}|\psi|^4$, 
we obtain  $E > E(K)$, where
\begin{equation}
E(K) = K + \frac{\omega^2N^2}{4K} + \frac{\lambda_2}{2} C_1 N K +
\frac{\lambda_3}{3} C_2 K^2 N.
\label{EK}
\end{equation}
When $\lambda_3 = 0$ we get the equation derived in \cite{Pit1}. 
Equation (\ref{EK}) should be compared with the corresponding 
variational expression (\ref{EvarND}), where 
the kinetic energy is given by $K = N\hbar\omega/(2\alpha^2)$ and 
$\alpha$ is the width of the cloud. As we see, the expression for the
energy (\ref{EK}) is very similar to the obtained by the variational
approach. However, (\ref{EK}) is valid for arbitrary time and
describes the nonstationary dynamics.
By using the variational expression (upper limit) for the
ground-state, and the right-hand-side of eq. (\ref{EK}) (lower limit), 
we can approach analytically the exact solution for the total energy
\begin{equation} 
E(K) < E < E_{var}.
\end{equation} 
For a more deep insight to the problem of stability, we need to obtain
the values of the constants $C_1$ and $C_2$.
This problem requires a generalization of the method suggested by
Weinstein in \cite{weinstein}, to be considered in a future work.

%%%%%%%%%%%%%%%%%%%%%%%%%%%%%% FIG. 5 %%%%%%%%%%%%%%%%%%%%%%%%%%%%%%%%%%%%
\begin{figure}[tbp]
\setlength{\epsfxsize}{0.9\hsize}
\setlength{\epsfysize}{0.8\hsize} 
\centerline{\epsfbox{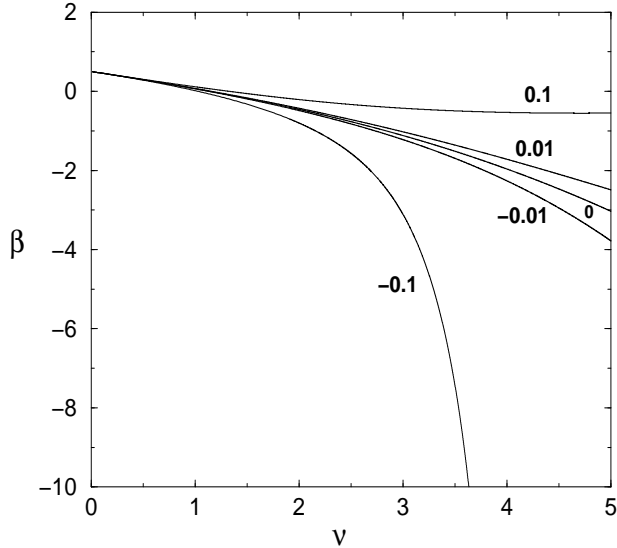}} 
\vskip 0.5cm
\caption{
Variational solutions for the chemical potential
(in dimensionless units) as functions of $\nu$, for $D=1$ and
different values of the three-body parameter $G_3$.} 
\end{figure}
%%%%%%%%%%%%%%%%%%%%%%%%%%%%%%        %%%%%%%%%%%%%%%%%%%%%%%%%%%%%%%%%%%%%%
%%%%%%%%%%%%%%%%%%%%%%%%%%%%%% FIG. 6 %%%%%%%%%%%%%%%%%%%%%%%%%%%%%%%%%%%%
\begin{figure}[tbp]
\setlength{\epsfxsize}{0.9\hsize} 
\setlength{\epsfysize}{0.8\hsize} 
\centerline{\epsfbox{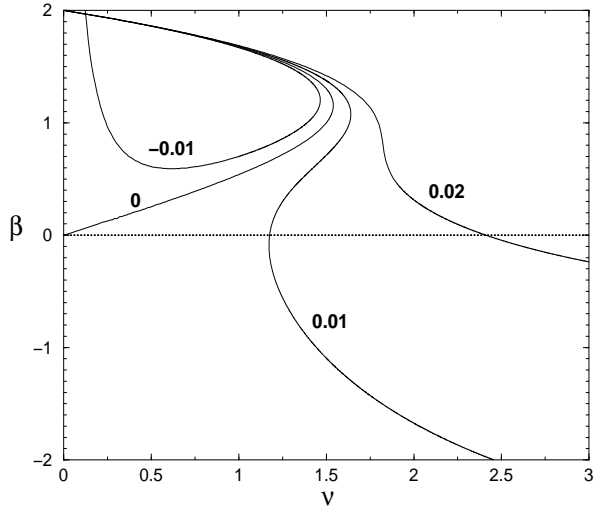}} 
\vskip 0.3cm
\caption{
The same as in Fig. 5, for $D=4$.}
\end{figure}
%%%%%%%%%%%%%%%%%%%%%%%%%%%%%%        %%%%%%%%%%%%%%%%%%%%%%%%%%%%%%%%%%%%%%

We should observe that exact numerical results, when $G_3=0$, have already
been considered in refs. \cite{rup95} (for $D=1$ and $D=3$),
in \cite{GFT} (for $D=$ 3), and in \cite{2D} (for $D=$2). 
 In \cite{ADV,GFT,GFTC}, for $D=3$, 
it was also considered the case with $G_3\ne 0$, and shown a kind
of first-order phase-transition in the condensate. In the present work,
we have extended the variational formalism, in case $G_3\ne 0$, for an
arbitrary $D-$dimension. In the following Figs. 5, 6 and 7, we present
our results for the chemical potential as a function of $\nu$, for a
set of given values of $G_3$, in case of $D =$ 1, 4 and 5. As one 
can observe in Fig. 5, even in case of $D=1$ one can reach a critical 
maximum limit for $\nu$, when $G_3$ is enough negative.
For $D =$ 4 and 5 (Figs. 6 and 7), we observe similar picture of
first-order phase-transition occurring for some specific values of $G_3$.

%%%%%%%%%%%%%%%%%%%%%%%%%%%%%% FIG. 7 %%%%%%%%%%%%%%%%%%%%%%%%%%%%%%%%%%%%
\begin{figure}[tbp]
\setlength{\epsfxsize}{0.9\hsize} 
\setlength{\epsfysize}{0.8\hsize} 
\centerline{\epsfbox{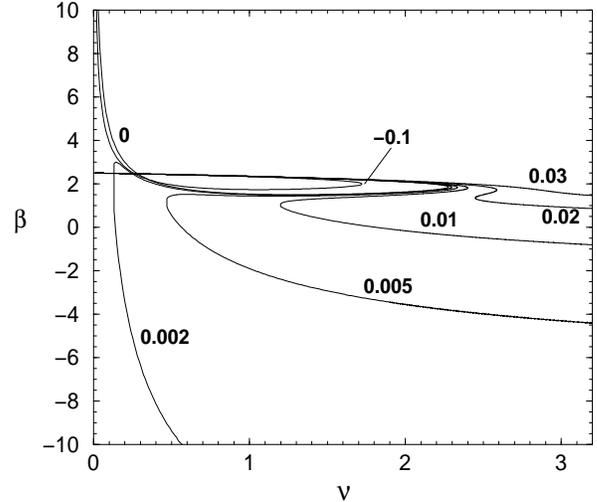}} 
\caption{
The same as in Fig. 5, for $D=5$.}
\end{figure}
%%%%%%%%%%%%%%%%%%%%%%%%%%%%%%        %%%%%%%%%%%%%%%%%%%%%%%%%%%%%%%%%%%%%%

In conclusion, in the present work, we first studied the 
stability and the critical number of atoms in arbitrary $D-$dimensions
using a variational procedure, for the case we have two-body
(attractive) and three-body contributions.
This part extends a previous analysis done in refs.~\cite{TW,Wad2}.
Next, we considered in more detail the case $D=2$.. We 
compared the variational results with exact numerical calculations
for the chemical potential, total energy, mean-square-radius and density. 
Finally, we extended numerically the approach for $D=2$, including an
effective three-body interaction term.
We studied  the sensibility of the critical numbers with
respect to corrections in the non-linear interaction.
The effective interaction considered in the equation contains a
trapped harmonic interaction, and two nonlinear terms, proportional to the
density $|\psi|^2$ (due to first-order two-body interaction) and to 
$|\psi|^4$ (due to first-order three-body interaction). 
We also verified, by a variational procedure, that a critical number of
particles exists only for $D\ge 2$, when the nonlinear term of the 
NLSE contains just the cubic term. In case of $D=1$, a critical
maximum number of atoms can exist with the addition of a negative
quintic term (three-body) in the NLSE. In all cases where the number of
atoms is limited, we
observed that the addition of a positive $|\psi|^4$ allows 
stable solutions beyond the critical number.
We also introduced an analysis of the collapse conditions, using ``the
virial theorem" approach given in \cite{Pit1}. 
The dynamics of the collapse was discussed in terms of the techniques
developed in \cite{weinstein}. In particular, we showed how the
exact energy can be approached in the case of $D=2$ with two and
three-body term contributions.

{\bf Acknowledgments} We are grateful to Jordan M. Gerton for the
suggestions and careful reading of the manuscript. 
This work was partially supported by Funda\c c\~ao
de Amparo \`a Pesquisa do Estado de S\~ao Paulo and Conselho Nacional de
Desenvolvimento Cient\'\i fico e Tecnol\'ogico.

\end{multicols}
\end{document}